\documentclass[runningheads]{llncs}
\usepackage{graphicx}
\usepackage{bm}
\usepackage{array}
\usepackage{amsmath}
\usepackage[colorlinks=true,linkcolor=blue,citecolor=blue,urlcolor=blue,]{hyperref}
\usepackage[doipre={doi:~}]{uri}
\usepackage{framed,multirow}
\usepackage{booktabs}
\usepackage{marvosym}
\usepackage{float}
\usepackage{color}
\begin{document}
\title{A Novel Hybrid Convolutional Neural Network for Accurate Organ Segmentation in 3D Head and Neck CT Images}
\titlerunning{Hybrid convolutional neural network}
%
\author{Zijie Chen\inst{*1,2,3,4} \and Cheng Li\inst{*5} \and Junjun He\inst{*1,2,6,7} \and Jin Ye\inst{1,2} \and Diping Song \inst{1,2} \and Shanshan Wang \inst{5,8,9} \and Lixu Gu \inst{6,7} \and Yu Qiao\inst{1,2}\textsuperscript{(\Letter)}}

\authorrunning{Z. Chen et al.}
%
\institute{Shenzhen Key Lab of Computer Vision and Pattern Recognition, SIAT-SenseTime Joint Lab, Shenzhen Institute of Advanced Technology, Chinese Academy of Sciences, Shenzhen, Guangdong, China \\ \email{yu.qiao@siat.ac.cn} \\
\and Shanghai AI Lab, Shanghai, China \\
\and Shenzhen Yino Intelligence Techonology Co., Ltd., Shenzhen, Guangdong, China
\and Shenying Medical Technology (Shenzhen) Co., Ltd., Shenzhen, Guangdong, China \\
\and Paul C. Lauterbur Research Center for Biomedical Imaging, Shenzhen Institute of Advanced Technology, Chinese Academy of Sciences, Shenzhen, Guangdong, China \\
\and School of Biomedical Engineering, Shanghai Jiao Tong University, Shanghai, China \\
\and Institute of Medical Robotics, Shanghai Jiao Tong University, Shanghai, China\\
\and Peng Cheng Laboratory, Shenzhen, Guangdong, China \\
\and Pazhou Lab, Guangzhou, Guangdong, China}
\maketitle              
\renewcommand{\thefootnote}{}
\footnotetext{* These authors contributed equally to this work.}
\renewcommand{\thefootnote}{\arabic{footnote}}
\begin{abstract}
Radiation therapy (RT) is widely employed in the clinic for the treatment of head and neck (HaN) cancers. An essential step of RT planning is the accurate segmentation of various organs-at-risks (OARs) in HaN CT images. Nevertheless, segmenting OARs manually is time-consuming, tedious, and error-prone considering that typical HaN CT images contain tens to hundreds of slices. Automated segmentation algorithms are urgently required. Recently, convolutional neural networks (CNNs) have been extensively investigated on this task. Particularly, 3D CNNs are frequently adopted to process 3D HaN CT images. There are two issues with naïve 3D CNNs. First, the depth resolution of 3D CT images is usually several times lower than the in-plane resolution. Direct employment of 3D CNNs without distinguishing this difference can lead to the extraction of distorted image features and influence the final segmentation performance. Second, a severe class imbalance problem exists, and large organs can be orders of times larger than small organs. It is difficult to simultaneously achieve accurate segmentation for all the organs. To address these issues, we propose a novel hybrid CNN that fuses 2D and 3D convolutions to combat the different spatial resolutions and extract effective edge and semantic features from 3D HaN CT images. To accommodate large and small organs, our final model, named OrganNet2.5D, consists of only two instead of the classic four downsampling operations, and hybrid dilated convolutions are introduced to maintain the respective field. Experiments on the MICCAI 2015 challenge dataset demonstrate that OrganNet2.5D achieves promising performance compared to state-of-the-art methods.

\end{abstract}
\section{Introduction}
Head and neck (HaN) cancers, such as oral cavity and nasopharynx, are one of the most prevalent cancer types worldwide \cite{Torre2015}. Treatment of HaN cancers relies primarily on radiation therapy. To prevent possible post-treatment complications, accurate segmentation of organs-at-risks (OARs) is vital during the treatment planning \cite{Han2008}. In the clinic, computed tomography (CT)-based treatment planning is routinely conducted because of its high efficiency, high spatial resolution, and the ability to provide relative electron density information. Manual delineation of OARs in CT images is still the primary choice regardless of the time-consuming and tedious process. Several hours are required to process the images of only one patient \cite{Harari2010}. Besides, it subjects to high inter- and intra-observer variations, which can significantly influence the prognosis of the treatment \cite{Brouwer2012}. Automatic segmentation methods are in urgent need to speed up the process and achieve robust outcomes. 

The low contrast of soft tissues in HaN CT images and the large volume size variations of different organs make it challenging to achieve automatic and accurate segmentation of all OARs in an end-to-end fashion. Conventional learning approaches often rely on one or multiple atlases or require the extraction of hand-crafted image features \cite{Chen2012,Wang2018}, which is difficult to be enough comprehensive and distinctive for the segmentation task. Deep neural networks, especially convolutional neural networks (CNNs), have proved to be highly effective for medical image segmentation in different applications \cite{Li2019,Qi2019}. Many efforts have been devoted to CNN-based segmentation of OARs in HaN CT images. To deal with the class imbalance issue caused by the differently sized organs, image patches based on certain prior knowledge were extracted before conducting CNN-based segmentation \cite{Ibragimov2017,Ren2018}. Two-step CNNs consisting of a region detector and a segmentation unit were also employed \cite{Men2019,Tang2019}. To make full use of the image information, a joint localization and segmentation network with a multi-view spatial aggregation framework was proposed \cite{Liang2020}. The inputs to these models were either 3D image patches lacking the global features or 2D images without the depthwise information. AnatomyNet was designed to specifically process whole-volume 3D HaN CT images \cite{Zhu2019}. The major contributions of AnatomyNet include a novel network architecture for effective feature extraction and a combined loss function to combat the class imbalance problem. Following AnatomyNet, FocusNet was proposed to better handle the segmentation of both large and small organs with a delicate network structure design \cite{Gao2019}.

Despite the inspiring results achieved, several issues exist in the developed approaches. First, some studies dealt with only 2D inputs and thus, did not fully exploit the 3D image information \cite{Ibragimov2017,Liang2020,Men2019}. Others conducted 3D convolutions but without paying attention to the different in-plane and depth resolutions \cite{Gao2019,Ren2018,Zhu2019}. The in-plane resolution of 3D HaN CT images is normally several times higher than the depth resolution. The direct employment of 3D convolutions can probably lead to the extraction of distorted image features, which might not be optimal for the segmentation task. Anisotropic convolutions have been proposed to solve this issue but without distinguishing the low-level and high-level features \cite{Liu2018}. Second, for networks processing whole volume 3D CT images (AnatomyNet and FocusNet), only one downsampling layer was used to preserve the information of small anatomies. Consequently, the receptive fields of these networks are limited. To increase the receptive field, DenseASPP with four dilation rates (3, 6, 12, and 18) was introduced to FocusNet \cite{Gao2019}. However, when the dilation rates of cascaded dilated convolutions have a common factor relationship, the gridding issue may appear that influence the segmentation accuracy \cite{Wang2018DilatedConv}. Besides, pure 3D networks also suffer from increased parameters and computational burden issues, which also limit the network depth and performance.

To address these issues, a hybrid convolutional neural network, OrganNet2.5D, is proposed in this work to improve the segmentation performance of OARs in HaN CT images. OrganNet2.5D integrates 2D convolutions with 3D convolutions to simultaneously extract clear low-level edge features and rich high-level semantic features. The hybrid dilated convolution (HDC) module is introduced to OrganNet2.5D as a replacement for the DenseASPP in FocusNet. HDC module is able to increase the network receptive field without decreasing the image resolutions and at the same time, avoid the gridding issue. OrganNet2.5D has three blocks: the 2D convolution block for the extraction of clear edge image features, the coarse 3D convolution block for the extraction of coarse high-level semantic features with a limited receptive field, and the fine 3D convolution block for the extraction of refined high-level semantic features with an enlarged receptive field through the utilization of HDC. Similar to AnatomyNet and FocusNet \cite{Gao2019,Zhu2019}, a combined loss of Dice loss and focal loss is employed to handle the class imbalance problem. The effectiveness of the proposed OrganNet2.5D is evaluated on two datasets. On the publicly available MICCAI Head and Neck Auto Segmentation Challenge 2015 dataset (MICCAI 2015 challenge dataset), promising performance is achieved by OrganNet2.5D compared to state-of-the-art approaches.

\section{Method}
\subsection{Dataset}
We evaluate the performance of our proposed model on two datasets. The first dataset is collected from two resources of 3D HaN CT images (the Head-Neck Cetuximab collection (46 samples) \cite{Clark2013} and the Martin Vallières of the Medical Physics Unit, McGill University, Montreal, Canada (261 samples)\footnote[1]{\url{https://wiki.cancerimagingarchive.net/display/Public/Head-Neck-PET-CT}}). This first dataset is utilized to validate the effectiveness of the different blocks of our model. Segmentation annotations of 24 OARs are provided by experienced radiologists with quality control management. We randomly grouped the 307 samples into a training set of 240 samples, a validation set of 20 samples, and a test set of 47 samples. To compare the performance of our proposed method to the existing approaches, we utilize the MICCAI 2015 challenge dataset \cite{Raudaschl2017}. There are 48 samples, among which 33 samples are provided as the training set, 10 as the offset test set, and the remaining 5 as the onsite test set. Manual segmentation of 9 OARs is available for the 33 training samples and 10 offset test samples. Similar to previous studies, we optimize our model with the training samples and report the model performance on the 10 offset test samples.

\begin{figure}[tb]
\centering
\includegraphics[width=12cm]{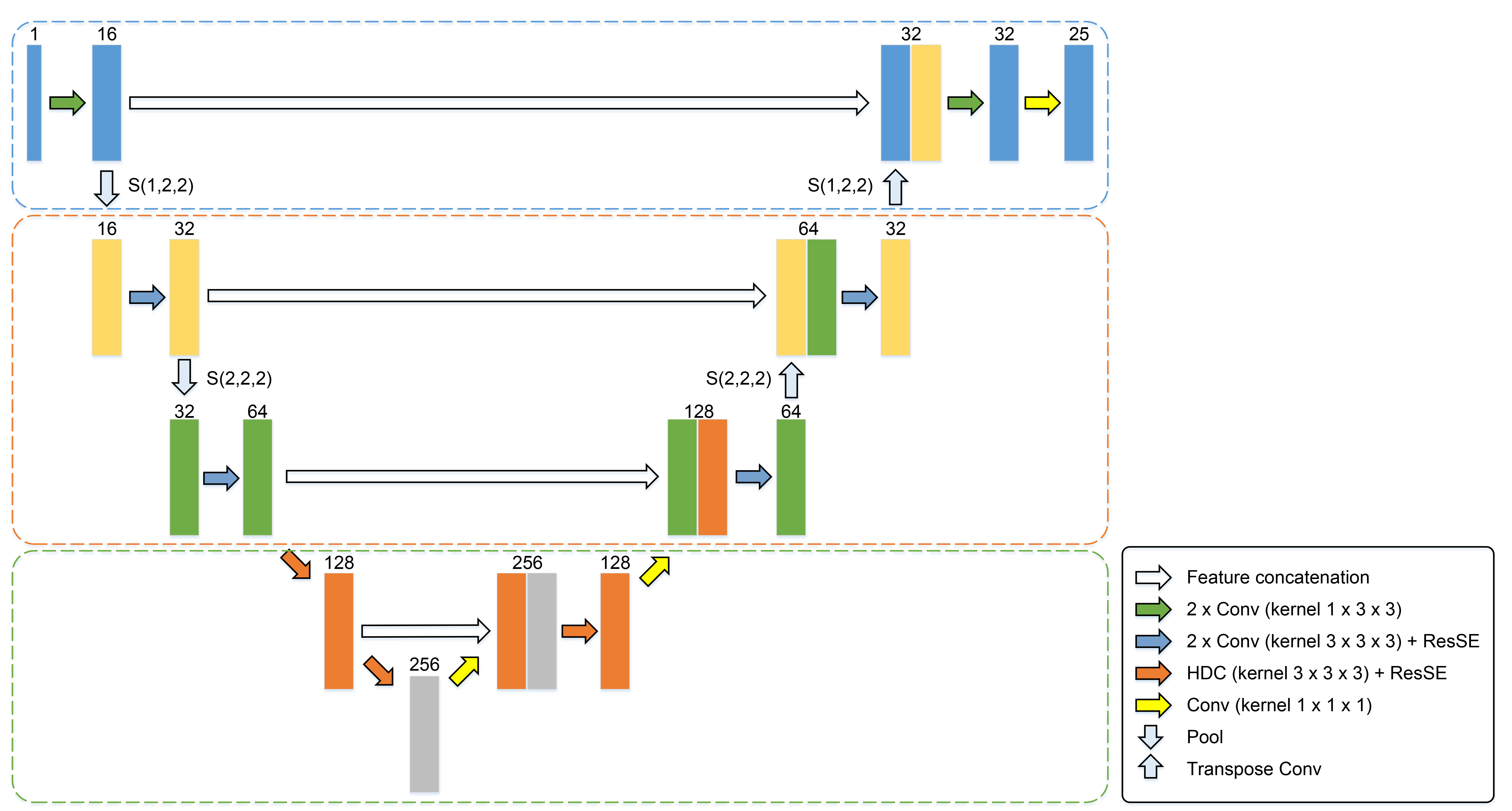}
\caption{An illustration of the proposed OrganNet2.5D network architecture. The blue, yellow, and green boxes indicate the 2D convolution block, the coarse 3D convolution block, and the fine 3D convolution block, respectively.}
\label{fig1}
\end{figure}

\begin{figure}[tb]
\centering
\includegraphics[width=12cm]{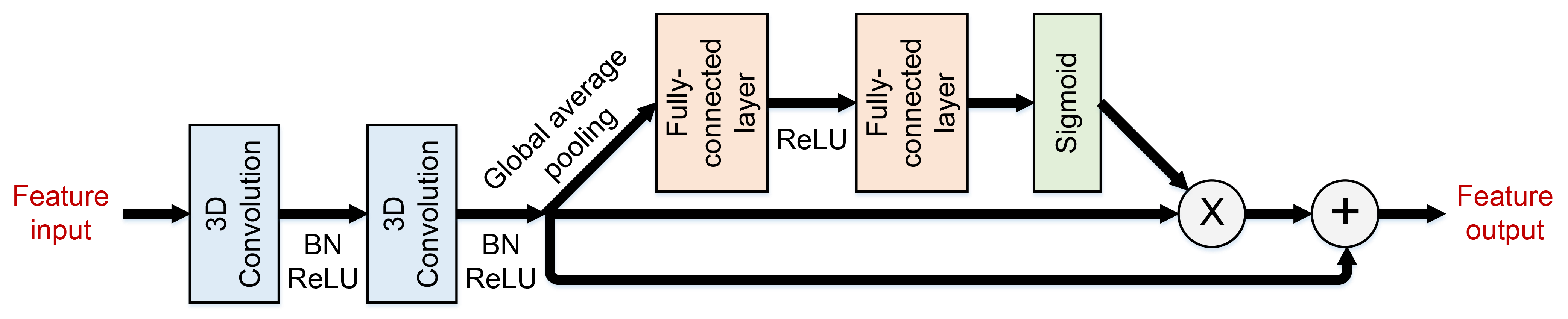}
\caption{The 2 $\times$ Conv + ResSE unit. ``X'' refers to element-wise multiplication and ``+'' is element-wise summation.}
\label{fig2}
\end{figure}

\subsection{Network architecture}
The overall network architecture of our proposed OrganNet2.5D is shown in Fig. \ref{fig1}. OrganNet2.5D follows the classic encoder-decoder segmentation network structure \cite{Ronneberger2015}. The inputs to our network are the whole volume 3D HaN CT images and the outputs are the segmentation results of the 25 categories for the first dataset (24 OARs and background) or 10 categories (9 OARs and background) for the MICCAI 2015 challenge dataset. OrganNet2.5D contains three major blocks, the 2D convolution block, the coarse 3D convolution block, and the fine 3D convolution block.

\subsubsection{2D convolution block.} The 2D convolution block is designed for the extraction of clear edge image features. It is widely accepted that during image encoding, the low-level features extract the geometric information and the high-level features extract the semantic information. Therefore, in our model, only the first two convolutions near the inputs and the corresponding last two convolutions near the outputs are replaced with 2D convolutions. Without the direct application of 3D convolutions, distorted image edge feature extraction can be avoided. Meanwhile, considering the different in-plane and depth image resolutions, in-plane downsampling is conducted with the 2D convolution block to calibrate the image features for the following 3D convolution operations.

\subsubsection{Coarse 3D convolution block.} The 2D convolution block is followed by the coarse 3D convolution block. To prevent information loss, especially for the small anatomies, only one downsampling is preserved. The coarse 3D convolution block is designed to extract rich semantic features that are important for the pixel-wise distinction task. Following the successful practice of existing methods, the basic unit of our coarse 3D convolution block is composed of two standard 3D convolution layers and one squeeze-and-excitation residual module (ResSE module, Fig. \ref{fig2}). The ResSE module is responsible for feature filtering to highlight the important features and suppress the irrelevant ones. With the filtered image features, the final segmentation step can concentrate more on the important features and better results can be expected.

\subsubsection{Fine 3D convolution block.} With the 2D convolution block and coarse 3D convolution block, clear edge and rich semantic image features are extracted. However, since only two downsampling layers are used (one 2D downsampling and one 3D downsampling), the receptive field of the network is limited. Without the global image information, the segmentation accuracy may be compromised. As such, a series of hybrid dilated convolution (HDC) modules is employed to integrate the global image information with the semantic features and at the same time, to prevent the gridding issue \cite{Wang2018DilatedConv}. Moreover, by using different dilation rates, multi-scale image features are extracted, which can better process the OARs of different sizes.

\subsection{Loss function}
A combination of focal loss and Dice loss is employed to prevent the model from biasing the large objects.

Focal loss forces the network to focus on the hard samples, which refers to the samples predicted by the network with high uncertainty. It is improved from the cross-entropy loss with both fixed and dynamic loss weighting strategies. The focal loss is calculated as:
\begin{equation}
L_{focal}= -\frac{1}{N}\sum_{n=1}^N\sum_{c=1}^C\alpha_c(1-p_n^c)^\gamma y_n^clogp_n^c
\label{eq01} 
\end{equation}
where $N$ refers to the sample size, $C$ refers to the different categories (25 for the first dataset and 10 for the second), $\alpha_c$ is the fixed loss weight of the $c^{th}$ OAR, $p\in[0,1]$ is the network prediction, $(1-p_n^c)^\gamma$ is the dynamic loss weight, and $y\in\{0,1\}$ is the manual label.

Dice loss deals with the class imbalance problem by minimizing the distribution distance between the network prediction and the manual segmentation. For multi-class segmentation, one Dice loss should be calculated for each class and the final Dice loss is the average over all the classes. In this work, the average Dice loss is calculated as:
\begin{equation}
L_{avgdice}= 1-\frac{1}{C}\sum_{c=1}^C\sum_{n=1}^N\frac{2\times p_n^c\times y_n^c}{p_n^c+y_n^c}
\label{eq02} 
\end{equation}

The final loss function for our network training is a weighted summation of the two losses:
\begin{equation}
L= L_{focal}+\lambda L_{avgdice}
\label{eq03} 
\end{equation}

For our experiments, we empirically set $\gamma=2$ and $\lambda = 1.0$. The fixed weights $\alpha_c$ in the focal loss for the first dataset are 0.5, 1.0, 1.0, 1.0, 4.0, 4.0, 4.0, 4.0, 4.0, 1.0, 1.0, 4.0, 1.0, 1.0, 3.0, 3.0, 1.0, 1.0, 1.0, 1.0, 1.0, 1.0, 3.0, 1.0, and 1.0 for the 25 categories (background, brain stem, eye left, eye right, lens left, lens right, optic nerve left, optic nerve right, optic chiasma, temporal lobes left, temporal lobes right, pituitary, parotid gland left, parotid gland right, inner ear left, inner ear right, middle ear left, middle ear right, tongue, temporomandibular joint left, temporomandibular joint right, spinal cord, mandible left, and mandible right), and for the second dataset are 0.5, 1.0, 4.0, 1.0, 4.0, 4.0, 1.0, 1.0, 3.0, and 3.0 for the 10 categories (background, brain stem, optic chiasma, mandible, optic nerve left, optic nerve right, parotid gland left, parotid gland right, submandibular left, submandibular right.)

\subsection{Implementation details}
All our models are implemented with PyTorch on an NVIDIA GeForce GTX 1080Ti GPU (11G) with a batch size of 2. The inputs to the networks are resized to $256\times256\times48$. Adam optimizer is utilized to train the models. The step decay learning rate strategy is used with an initial learning rate of 0.001 that is reduced by a factor of 10 every 50 epochs until it reaches 0.00001. Two evaluation metrics are calculated to characterize the network performance, the Dice score coefficient (DSC) and the 95\% Hausdorff distance (95HD).

\section{Experimental results}
\subsection{Results on the collected public dataset}
Ablation studies regarding our network design are conducted. Average DSC and 95HD on the test set are listed in Table \ref{tab1}. DSC values of the 10 small organs are presented in Table \ref{tab2}. See supplementary material for results on all 24 organs. Four network configurations are involved. 3DUNet-SE refers to the baseline where 3D UNet is combined with the ResSE module. 3DUNet-SE includes only the coarse 3D convolution block in Fig. \ref{fig1}. Introducing the 2D convolution block to 3DUNet-SE, we obtain the 3DUNet-SE-2D model. 3DUNet-SE-2D-C replaces the HDC module in the proposed OrganNet2.5D (Fig. \ref{fig1}) with standard 3D convolutions, and 3DUNet-SE-2D-DC replaces the HDC module with dilated convolutions of the same dilation rate of 2.

Overall, our proposed model achieves the highest mean DSC and lowest mean 95HD. Statistical analysis confirms that our model performs significantly better than the other network configurations ($p<0.05$ with paired $t$-tests of the DSC values). These results reflect that both the 2D convolution block and the fine 3D convolution block can enhance the segmentation results. Furthermore, our proposed OrganNet2.5D gives excellent performance on small organ segmentation by generating the best results for 7 of the 10 small organs (Table \ref{tab2}).
\begin{table}[!t]
\caption{\label{tab1} Segmentation performance on the first dataset averaged over the 24 OARs with different network configurations}
\centering
\begin{tabular}{m{1.5cm}<{\centering}m{2cm}<{\centering}m{2cm}<{\centering}m{2cm}<{\centering}m{2cm}<{\centering}m{2cm}<{\centering}}
\toprule[1pt]
Models & 3DUNet-SE & 3DUNet-SE-2D & 3DUNet-SE-2D-C2 & 3DUNet-SE-2D-DC2 & Proposed \\
\midrule[1pt]
DSC (\%) & $83.9 \pm 2.0$  & $84.4 \pm 2.0$ & $84.2  \pm 2.1$ & $84.3 \pm 2.1$ & $\mathbf{84.6 \pm 1.9}$\\
95HD & 3.38 & 3.39 & 3.24 & 3.41 & \textbf{3.07}\\
\bottomrule[1pt]
\end{tabular}
\end{table}

\begin{table}[!t]
\caption{\label{tab2} DSC of 10 small organs on the first dataset with different network configurations (\%)}
\centering
\begin{tabular}{m{2cm}<{\centering}m{1.6cm}<{\centering}m{2cm}<{\centering}m{2cm}<{\centering}m{2cm}<{\centering}m{2cm}<{\centering}}
\toprule[1pt]
Models & 3DUNet-SE & 3DUNet-SE-2D & 3DUNet-SE-2D-C & 3DUNet-SE-2D-DC & Proposed \\
\midrule[1pt]
Lens L & $82.7 \pm7.8$ & $\mathbf{84.5 \pm 6.8}$ & $83.5 \pm7.5$ & $83.2 \pm 6.5$ & $84.4 \pm 6.6$\\
Lens R & $82.6 \pm 5.6$ & $\mathbf{84.1 \pm 5.8}$ & $83.3 \pm 6.9$ & $83.9 \pm 5.4$ & $83.9 \pm 5.3$\\
Opt. Ner. L & $70.9 \pm 10.0$ & $71.0 \pm 10.1$ & $70.7 \pm 10.2$ & $70.5 \pm 10.8$ & $\mathbf{71.1 \pm 10.3}$\\
Opt. Ner. R & $70.1 \pm 8.7$ & $71.7 \pm 8.9$ & $\mathbf{71.8 \pm 9.3}$ & $70.4 \pm 9.1$ & $71.2 \pm 9.2$\\
Opt. Chiasm & $57.2 \pm 14.3$ & $59.6 \pm 15.1$ & $57.8 \pm 15.4$ & $58.0 \pm 14.9$ & $\mathbf{59.8 \pm 14.8}$\\
Pituitary & $74.4 \pm 11.6$ & $75.1 \pm 13.0$ & $74.3 \pm 12.4$ & $75.1 \pm 11.4$ & $\mathbf{75.6 \pm 12.3}$\\
Mid. Ear L & $86.4 \pm 5.1$ & $86.6 \pm 4.8$ & $86.5 \pm 5.1$ & $86.7 \pm 4.9$ & $\mathbf{86.9 \pm 5.1}$\\
Mid. Ear R & $85.4 \pm 4.1$ & $85.5 \pm 4.3$ & $85.8 \pm 4.6$ & $85.6 \pm 4.7$ & $\mathbf{85.9 \pm 4.3}$\\
T.M.J. L & $83.5 \pm 7.2$ & $83.8 \pm 7.5$ & $83.8 \pm 7.1$ & $82.7 \pm 7.8$ & $\mathbf{83.8 \pm 7.2}$\\
T.M.J. R & $82.1 \pm 8.3$ & $81.6 \pm 8.8$ & $82.8 \pm 7.9$ & $82.7 \pm 8.2$ & $\mathbf{82.9 \pm 7.9}$\\
\bottomrule[1pt]
\end{tabular}
\end{table}

\begin{table}[!t]
\caption{\label{tab3} Segmentation results on the MICCAI 2015 challenge dataset}
\centering
\begin{tabular}{m{1.9cm}<{\centering}m{1.2cm}<{\centering}m{1.7cm}<{\centering}m{1.5cm}<{\centering}m{1.5cm}<{\centering}m{1.5cm}<{\centering}m{1.6cm}<{\centering}m{1.5cm}<{\centering}}
\toprule[1pt]
Models & MICCAI 2015 & AnatomyNet \cite{Zhu2019} & FocusNet \cite{Gao2019} & SOARS \cite{Guo2020} & SCAA \cite{Tang2021} & Proposed \\
\midrule[1pt]
Brain Stem & 88.0 & $86.7 \pm 2$ & $87.5 \pm 2.6$ & $87.6 \pm 2.8$ & $\mathbf{89.2 \pm 2.6}$ & $87.2 \pm 3.0$\\
Opt. Chiasm & 55.7 & $53.2 \pm 15$ & $59.6 \pm 18.1$ & $64.9 \pm 8.8$ & $62.0 \pm 16.9$ & $\mathbf{66.3 \pm 7.4}$\\
Mandible & 93.0 & $92.5 \pm 2$ & $93.5 \pm 1.9$ & $95.1 \pm 1.1$ & $\mathbf{95.2 \pm 1.3}$ & $92.2 \pm 2.1$\\
Opt. Ner. L & 64.4 & $72.1 \pm 6$ & $73.5 \pm 9.6$ & $75.3 \pm 7.1$ & $\mathbf{78.4 \pm 6.1}$ & $75.0 \pm 7.8$\\
Opt. Ner. R & 63.9 & $70.6 \pm 10$ & $74.4 \pm 7.2$ & $74.6 \pm 5.2$ & $\mathbf{76.0 \pm 7.5}$ & $74.1 \pm 5.1$\\
Parotid L & 82.7 & $88.1 \pm 2$ & $86.3 \pm 3.6$ & $88.2 \pm 3.2$ & $\mathbf{89.3 \pm 1.5}$ & $86.7 \pm 2.6$\\
Parotid R & 81.4 & $87.4 \pm 4$ & $87.9 \pm 3.1$ & $88.2 \pm 5.2$ & $\mathbf{89.2 \pm 2.3}$ & $85.8 \pm 4.9$\\
Subman. L & 72.3 & $81.4 \pm 4$ & $79.8 \pm 8.1$ & $\mathbf{84.2 \pm 7.3}$ & $83.2 \pm 4.9$ & $82.1 \pm 5.8$\\
Subman. R & 72.3 & $81.3 \pm 4$ & $80.1 \pm 6.1$ & $\mathbf{83.8 \pm 6.9}$ & $80.7 \pm 5.2$ & $82.1 \pm 4.1$\\
Mean DSC & 74.9 & 79.2 & 80.3 & 82.4 & \textbf{82.6} & 81.3\\
\bottomrule[1pt]
\end{tabular}
\end{table}

\subsection{Results on MICCAI 2015 challenge dataset}
We compare the performance of our proposed model to the state-of-the-art methods on the MICCAI 2015 challenge dataset (Table \ref{tab3}). It should be noted that in the table, the MICCAI 2015 results were the best results obtained for each OAR possibly by different methods. AnatomyNet was trained with additional samples except for the MICCAI 2015 challenge dataset. All the results of existing methods are adopted from the respective papers without method re-implementation to avoid implementation biases.


Segmentation results show that our proposed model achieves better performance than the three most prevalent methods in the field (MICCAI 2015, AnotomyNet, and FocusNet) indicated by the mean DSC, which confirms the effectiveness of the proposed network. Compared to the two recently published methods, SOARS and SCAA, our method is slightly worse. However, it should be noted that SOARS utilized neural network search to find the optimal network architecture \cite{Guo2020}, which is more computationally intensive. SCAA combined the 2D and 3D convolutions with a very complicated network design \cite{Tang2021}. Nevertheless, with the simple and easy-to-implement architecture, our OrganNet2.5D still performs the best when segmenting the smallest organ, optic chiasma. This observation reflects the suitability of our network modifications and training strategy for our task. Visual results lead to similar conclusions as to the quantitative results (See supplementary material for details).

\section{Conclusion}
In this study, we present a novel network, OrganNet2.5D, for the segmentation of OARs in 3D HaN CT images, which is a necessity for the treatment planning of radiation therapy for HaN cancers. To fully utilize the 3D image information, deal with the different in-plane and depth image resolutions, and solve the difficulty of simultaneous segmentation of large and small organs, OrganNet2.5D consists of a 2D convolution block to extract clear edge image features, a coarse 3D convolution block to obtain rich semantic features, and a fine 3D convolution block to generate global and multi-scale image features. The effectiveness of OrganNet2.5D was evaluated on two datasets. Promising performance was achieved by our proposed OrganNet2.5D compared to the state-of-the-art approaches, especially on the segmentation of small organs.

\subsubsection{Acknowledgements.} 
This research is partially supported by the National Key Research and Development Program of China (No. 2020YFC2004804 and 2016YFC0106200), the Scientific and Technical Innovation 2030-"New Generation Artificial Intelligence" Project (No. 2020AAA0104100 and 2020AAA0104105), the Shanghai Committee of Science and Technology, China (No. 20DZ1100800 and 21DZ1100100), Beijing Natural Science Foundation-Haidian Original Innovation Collaborative Fund (No. L192006), the funding from Institute of Medical Robotics of Shanghai Jiao Tong University, the 863 national research fund (No. 2015AA043203), Shenzhen Yino Intelligence Techonology Co., Ltd., Shenying Medical Technology (Shenzhen) Co., Ltd., the National Natural Science Foundation of China (No. 61871371 and 81830056),  the Key-Area Research and Development Program of GuangDong Province (No. 2018B010109009), the Basic Research Program of Shenzhen (No. JCYJ20180507182400762), and the Youth Innovation Promotion Association Program of Chinese Academy of Sciences (No. 2019351).

\end{document}